\documentstyle[12pt,epsfig]{article}
\newlength{\dinwidth}
\newlength{\dinmargin}
\newcommand{\ba}{\begin{array}}
\newcommand{\ea}{\end{array}}
\newcommand{\be}{\begin{equation}}
\newcommand{\ee}{\end{equation}}
\newcommand{\bea}{\begin{eqnarray}}
\newcommand{\eea}{\end{eqnarray}}

\newcommand{\lsim}{\mathrel{\raisebox{-.6ex}{$\stackrel{\textstyle<}{\sim}$}}}

\begin{document}
\thispagestyle{empty}
\addtocounter{page}{-1}
\begin{flushright}
SNUST-000901\\
{\tt hep-th/0012165}
\end{flushright}
\vspace*{1.3cm}
\centerline{\Large \bf Global Black $p$-Brane World:}
\vskip0.3cm
\centerline{\Large \bf A New Approach to Stable Mass Hierarchy 
~\footnote{Work supported in part by BK-21 Initiative in Physics (SNU - 
Project 2), KRF International Collaboration Grant 1998-010-192, KOSEF 
Interdisciplinary Research Grant 98-07-02-07-01-5, and KOSEF Leading 
Scientist Program 2000-1-11200-001-1 (S.-J. R.) and KOSEF Basic Research 
Program No. 2000-1-11200-001-3 (Y. K.).}}
\vspace*{1.4cm} 
\centerline{\bf Sei-Hoon Moon ${}^a$, Soo-Jong Rey ${}^{a}$ {\rm and}
Yoonbai Kim ${}^b$}
\vspace*{0.8cm}
\centerline{\it School of Physics \& Center for Theoretical Physics}
\vspace*{0.2cm}
\centerline{\it Seoul National University, Seoul 151-742 Korea ${}^a$}
\vspace*{0.6cm}
\centerline{\it Department of Physics \& Institute of Basic Science}
\vspace*{0.2cm}
\centerline{\it Sungkyunkwan University, Suwon 440-746 Korea ${}^b$}
\vspace*{1cm}
\centerline{\tt jeollo@phya.snu.ac.kr, \quad sjrey@gravity.snu.ac.kr, 
\qquad yoonbai@cosmos.skku.ac.kr}
\vspace*{1.5cm}
\centerline{\bf abstract}
\vspace*{0.5cm}
We find a class of extremal black hole-like global $p$-brane in 
higher-dimensional gravity with a negative cosmological constant. 
The region inside the $p$-brane horizon possesses all essential features 
required for the Randall-Sundrum-type brane world scenario.
The set-up allows to interpret the horizon size as the 
compactification size in that the Planck scale $M_{\rm Pl}$ is 
determined by the fundamental scale $M_*$ and the horizon size $r_{\rm H}$ 
via the familiar relation 
$M_{\rm Pl}^2\sim M_*^{2+n}r_{\rm H}^n$, and the gravity behaves as expected 
in a world with $n$-extra dimensions compactified with size $r_{\rm H}$. 
Most importantly, a stable mass hierarchy between $M_{\rm Pl}$ and $M_*$ can 
be generated from topological charge of the $p$-brane and the horizon 
size $r_{\rm H}$ therein. We also offer a new perspective on various 
issues associated to the brane world scenarios including the cosmological 
constant problem.

\vspace*{1.1cm}

\baselineskip=18pt

\newpage

\setcounter{equation}{0}

%%%%%%%%%%%%%%%%%%%%%%%%%%%%%%%%%%%%%%%%%%%%%%%%%%%%%%%%%%%%%%%%%%%%%%%%%%%
\section{Introduction}
%%%%%%%%%%%%%%%%%%%%%%%%%%%%%%%%%%%%%%%%%%%%%%%%%%%%%%%%%%%%%%%%%%%%%%%%%%%

In exploring brane world scenarios in recent years, seemingly two alternative
directions have been pursued. 

First direction, as pioneered by Arkani-Hamed, 
Dimopoulos and Dvali (ADD), involves reducing the fundamental mass scale from 
the four-dimensional Planck scale, $M_{\rm Pl} \sim10^{19}$GeV, 
to the electroweak scale, $m_{\rm EW}\sim10^{3}$GeV,
by introducing {\sl large but compact} extra dimensions. Current experimental
bound suggests that the extra dimensions could be as big as sub-millimeter 
scale when the extra dimensions are two \cite{ADD}. 
In the simple examples of \cite{ADD,STR1,STR2} proposal, 
the spacetime geometry is a direct product of four-dimensional Minkowski 
spacetime and a spatial $n$-torus of volume ${\rm Vol}_n \sim R^n$. Then, at 
low-energy, effective four-dimensional Planck scale $M_{\rm Pl}$ is set by the 
Gauss' relation, $M_{\rm Pl}^2\, \sim \, M_*^{2+n}R^n$, where $M_*$ refers to 
the fundamental scale set by the higher-dimensional gravity. The hierarchy 
between $M_{\rm Pl}$ and $M_*$ can be very large if $RM_*\gg1$. The ADD 
scenario is extremely attractive as not only the hierarchy problem is explained
but also rich phenomenology of collider physics \cite{GRW,coll,review} and 
cosmology \cite{cosmo} are anticipated. In the
near future, we may even be able to probe strong quantum gravity effects. On 
the other hand, as it stand now, the ADD scenario is afflicted by various 
unsolved fundamental problems. One of them is stabilization problem 
of the large extra dimensions or, equivalently, dynamical determination of 
large extra dimensions\footnote{For two extra dimensions, an interesting
proposal to this problem has been put forward in \cite{AB}}.
Paraphrasing, the central question posed by the ADD
scenario is \lq\lq Why are the extra dimensions so large?".

The second direction, as proposed by Randall and Sundrum (RS), 
is a scheme in which a five-dimensional gravity contains strongly 
gravitating three-branes, which induces a {\sl warped or nonfactorizable} 
spacetime geometry. In the first RS model \cite{RS1}, which consists of two
three-branes, a large mass hierarchy was achieved without large extra 
dimensions. In the second 
RS model \cite{RS2},  which consists only one three-brane of a positive 
tension, the extra dimension is noncompact yet the low-energy gravity on the 
three-brane still comes out to be effectively four-dimensional at low-energy,
up to power-like corrections at short distance. This has to do with the fact
that the five-dimensional graviton is decomposed into a massless bound state 
confined to the three-brane as well as a continuum of the 
Kaluza-Klein (KK) states. For low-energy processes on the three-brane, 
the bound state dominates over the KK states and gives rise to an 
inverse-square law provided the five-dimensional bulk curvature is sufficiently
large. The RS scenario has been extended to various directions. One extension 
involves turning on a cosmological constant on the brane \cite{bent,cosm}. 
Other extensions include higher dimensional embedding \cite{highd} and 
realizations of domain walls in gravity coupled to scalars 
\cite{gubs,gremm,CEHS,bs,ADDK}. Whether the RS scenario can be embedded into
a more complete setting has been addressed within the setup of supergravity 
\cite{sugra,nogo} and string theories \cite{string}.

In Refs.\cite{CK,Gre,Vil,GS,shap,Oda}, attempts were made to embed these two 
scenarios by utilizing {\sl gravitating defects} in higher dimensions.
Cohen and Kaplan have considered a brane world, which may be viewed 
as a global string in two extra dimensions without bulk cosmological constant 
\cite{CK}. Its spacetime geometry is strongly warped and, similar to the case
in $(2+1)$-dimensions, exhibits a curvature singularity at a finite distance
from the string core. They argued that, imposing a unitary boundary condition,
the location of the singularity may play a role of the boundary of the extra
dimensions.  
Gregory claimed \cite{Gre} that, in the presence of a negative bulk 
cosmological constant, there does not exist a non-singular global vortex 
solution which asymptotes to anti-de Sitter spacetime, and that a 
genuine non-singular global vortex can form a (physical) singularity-free
warped geometry provided the extra-dimensions asymptote to a cylinder.
Extensions to higher extra-dimensional defects (for example, a global monopole
in three extra-dimensions) and to inclusion of gauge fields were discussed 
in Ref.\cite{Vil,shap} and in Ref.\cite{GS,shap}, respectively. 
 
In this paper, we would like to present a new sort of 
brane-world scenario, which 
accommodates features of the above two scenarios and, most significantly, 
offers a {\sl topological mechanism} of stabilizing a large mass hierarchy. Our
setup will be, in a higher-dimensional anti-de Sitter spacetime background, 
a class of extremal black brane made of $p$-dimensional global defect, which 
shall be referred in what follows as \lq\lq global black $p$-brane".
We focus on the interior region inside the black $p$-brane horizon and study 
effective four-dimensional gravity at low energy on a `Standard Model'-brane,
which could be provided by the core brane or by 
a `probe' brane located near the 
core of the interior region. A crucial feature of the interior region is that, 
even though asymptoting to an infinitely long throat, its volume is finite.
As such, the `Standard Model'-brane will behave locally as a domain wall 
embedded in an anti-de Sitter space. The bulk graviton in this background 
yields a massless localized mode on the $p$-brane, reproducing the correct 
$(p+1)$-dimensional gravity at long distance, in the same way as in the RS 
scenario. In our new-brane world scenario, a noteworthy feature is 
that the $p$-brane horizon size is interpretable as a compactification size, 
as the Planck scale $M_{\rm Pl}$ is determined in units of
the fundamental scale $M_*$ and the horizon size $r_{\rm H}$ via:
\bea 
M_{\rm Pl}^{p-1}\quad \sim \quad M_*^{p-1+n} \, r_{\rm H}^n.
\nonumber
\eea
It also shows that, by varying the horizon size $r_{\rm H}$, the effective 
gravity on the `Standard Model'-brane interpolates between 
the RS scenario and the ADD scenario limits when more than one extra dimension
are present. 
A large mass hierarchy is then simply 
translated into large horizon size. The most interesting feature 
of our scenario is that the horizon size $r_{\rm H}$, measured
in units of the fundamental scale $M_*$, is determined by the topological 
charge $Q$ of the global $p$-brane. For instance, for $n=2$ and $p=3$, 
\bea
r_{\rm H}^2 = {16 \pi v^2 \over M_*^4 \vert \Lambda \vert} Q^2
\qquad \rightarrow \qquad
r_{\rm H} \sim {1 \over M_*} \vert Q \vert. 
\nonumber
\eea
A large mass hierarchy then follows by taking $\vert Q \vert \gg 1$.
As a consequence, in our new brane-world scenario, the macroscopic 
compactification size is derived explicitly from a microscopic theory 
equipped with a fundamental scale and stabilization of the large hierarchy 
is ensured by the discrete and conserved nature of the $p$-brane 
topological charge $Q$.

%%%%%%%%%%%%%%%%%%%%%%%%%%%%%%%%%%%%%%%%%%%%%%%%%%%%%%%%%%%%%%%%%%%%%%%%%
\section{Global Black $p$-Brane Solutions}
%%%%%%%%%%%%%%%%%%%%%%%%%%%%%%%%%%%%%%%%%%%%%%%%%%%%%%%%%%%%%%%%%%%%%%%%%%%

Our starting point is a $D$-dimensional Einstein gravity coupled to a field 
theory of an $n$-component scalar ${\bf \Phi}^a \, (a = 1, \cdots, n)$ 
($n$-vector model). We assume that the $O(n)$ global symmetry is spontaneously broken to $O(n-1)$ and, at the ground-state $\langle {\bf \Phi}^2 \rangle
= v^2$, the vacuum energy density is negative. Dynamics of the theory is 
governed by the following action:
\begin{equation}\label{th}
S=\int dx^D\sqrt{g_{D}}\left[-\frac{M^{D-2}_*}{16\pi}(R+2\Lambda)-
\frac12 g^{AB} \nabla_A{\bf \Phi} \cdot \nabla_B {\bf \Phi} - 
\frac{\lambda}{4} ({\bf \Phi}^2 -v^2)^2 \right],
\end{equation}
where $A, B, \cdots$ denote $D$-dimensional spacetime indices and $M_*$ 
sets a `fundamental mass scale' of the theory. This theory admits 
$p$-dimensional topological solitons, where $(p+1) = D - n$. We shall refer as 
`global $p$-brane' in what follows. Note that ${\bf \Phi}^a$ and $v$ have a 
mass dimension of $(p+n-1)/2$ and the negative bulk cosmological constant 
($\Lambda<0$) has a mass dimension of $2$. 
We shall use the notation $\{x^\mu\}$ with $\mu=0,...,p$ for the coordinates 
on the $p$-brane worldvolume, and $\{{\bf y}^a\}$ with $a=1,..., n$ for the 
coordinates in the extra dimensions and are related to the spherical 
coordinates of ${\bf S}_{n-1}$ by the usual relations: 
$y^a= (r\cos\theta_1,\cdots\cdots,
r\sin\theta_1 \cdots \cos\theta_{d-1}, r\sin\theta_1\cdots\sin\theta_{d-1})$ 
with $r^2={\bf y} \cdot {\bf y}$.
 
The global $p$-brane is defined by a topologically nontrivial 
mapping of the vacuum manifold 
$O(n)/O(n-1)$ to the extra dimensions ${\bf S}_{n-1}$. For a unit topological 
charge, we take ans\"atz for the scalar field configuration as:  
\begin{equation}\label{sconf}
{\bf \Phi}^a=vf(r)\frac{{\bf y}^a}{r}.
\end{equation}
Here, for regularity of the $p$-brane, the spherically symmetric function 
$f(r)$ is required to satisfy $f(r)=0$ at the center and approach 
$f(r) = 1$ outside core. For simplicity, in this paper, we shall take a version
of `thin-wall approximation' so that only the region outside the core 
\footnote{Detailed analysis without assumption $f(r)=1$ will be treated in 
Ref.\cite{Moon}.} is considered. We take the 
following metric ans\"atz of Schwarzschild-type:  
\begin{equation}\label{metric1}
ds^{ 2}=e^{2N(r)}B(r) \hat{g}_{\mu\nu}(x)dx^\mu dx^\nu+\frac{d^{2}r}{B(r)}
+r^{2}d\Omega^{2}_{n-1},
\end{equation}
where $\hat{g}_{\mu\nu}(x)$ denotes the metric of $(p+1)$-dimensional 
subspace longitudinal to the global $p$-brane worldvolume.

With the ans\"{a}tz Eqs.(\ref{sconf}) and (\ref{metric1}), one obtains, from 
the action Eq.(\ref{th}), the following set of field equations:
\begin{eqnarray}
(p+1)BN'^2+\frac{2p+3}{2}B'N'
+\frac{p}{4}\frac{B'^2}{B}+BN''+\frac12B''
+\frac{n-1}{r}\left(BN'+\frac12B'\right) \nonumber\\ 
+\frac{1}{p+1}\frac{\hat{R}}{e^{2A}B}=-\frac{2}{p+n-1}\Lambda, \label{ein1}\\
(p+1)\left(BN'^2+\frac32B'N'+BN''+\frac12B''\right)+\frac{n-1}{2}\frac{B'}{r}
=-\frac{2}{p+n-1}\Lambda,\label{ein3}\\
(p+1)\frac{BN'}{r}+\frac{p+2}{2}\frac{B'}{r}+(n-2)\frac{B-1}{r^2}
=-\frac{8\pi G_D}{r^2} -\frac{2}{p+n-1}\Lambda\label{ein4},
\end{eqnarray}
where we have rescaled both the coordinates and the cosmological constant
as $x_A(\equiv \sqrt{\lambda}v~x_A)$, $\Lambda(\equiv\Lambda/\lambda v^{2})$,
respectively. We have also defined $G_D\equiv v^{2}/M_{\ast}^{D-2}$ and
$\hat{R}$ for the curvature associated with the metric $\hat{g}_{\mu\nu}$. 

The field equations Eqs.(\ref{ein1}-\ref{ein4}) being highly nonlinear, 
the spacetime around the global $p$-brane may develop singularities. 
As discussed in Ref.\cite{Vil}, global structure of the spacetime 
depends largely on the number of extra dimensions $n$, the bulk cosmological 
constant $\Lambda$, and the $p$-brane worldvolume curvature $\hat{R}$. 
If $\Lambda = 0 $, for $n=2$, the spacetime has a singularity, much as in the
cosmic string background in four dimensions \cite{CK}, while, for $n\geq3$, 
the spacetime is regular and corresponds to that of the global monopole. 
In case $\Lambda < 0$, the cosmological constant acts as a censor against 
the singularity, depending on the value of the worldvolume curvature ${\hat R}$.
If ${\hat R} = 0$, the spacetime is always regular no matter how small the 
cosmological constant may be.  If ${\hat R} \ne 0$, the spacetime may become 
singular at a finite proper distance from the global $p$-brane core. 
For the black brane solutions, the case with $\hat{R}=0$ corresponds to an 
extremal black $p$-brane, while the case with $\hat{R}<0$ to a non-extremal 
black $p$-brane. What we expect is that, as in the case of the supergravity 
black $p$-brane, a singularity may be developed at the inner horizon of a 
non-extremal global $p$-brane \cite{Moon}. For the most part of this paper, 
we will take that the $p$-brane is extremal and hence has a Ricci flat 
worldvolume, viz. $\hat{R}=0$.

Finding an exact analytic solution to the field equations 
Eq.(\ref{ein1})-(\ref{ein4}) being impossible, let us try to examine behavior 
of the global $p$-brane at various regions. Outside the core, the spacetime is 
determined solely by the relative magnitude of the scalar field energy 
density ($8\pi G_D/r^2$) to the cosmological constant ($|\Lambda|$). 
Analysis via series expansion and numerical integration indicates that the 
product $e^{2N}B$, which determines the $(00)$-th component of the spacetime 
metric, decreases monotonically near the core for $2\pi G_D>|\Lambda|$, and can 
be made to vanish at a finite distance by tuning $G_D$.
  
In the region far from the center where 
$8\pi G_D/r^2\ll|\Lambda|$, the aforementioned metric functions behave as:
$
N(r)\sim N_{\infty}~{\rm and}~B(r)\sim B_{\infty}r^2,
$
where $N_{\infty}$ is a calculable constant (determined by matching the 
function to the interior-region) and $B_\infty\equiv 2|\Lambda|/(p+n)(p+n-1)$. 
We thus find that the exterior-region asymptotes to a $D$-dimensional anti-de 
Sitter spacetime ($AdS_D$). By introducing a proper radial distance
 $\chi\equiv\int dr/\sqrt{B(r)}$, the spacetime in the exterior-region can 
be expressed as follows:
\begin{eqnarray}\label{amet}
ds^2\, := \, e^{2\sqrt{B_\infty}\chi}~\bar{g}_{\mu\nu}dx^\mu dx^\nu+d\chi^2
             +e^{2\sqrt{B_\infty}\chi}~d\Omega_{n-1}^2.
\end{eqnarray}
Here, $\bar{g}_{\mu\nu}(x)$ is a general Ricci-flat metric on the brane,
satisfying $(p+1)$-dimensional vacuum Einstein field equation: 
$\bar{R}_{\mu\nu}(\bar{g})=0$. 

The global $p$-brane horizon is where the timelike Killing vector $\partial_t$
becomes null and is formed in a region where the energy 
density of the scalar field and the bulk cosmological constant are comparable. 
Thus, we will take that there exists a horizon at a {\sl finite} coordinate 
distance $r=r_{\rm H}$ from the $p$-brane center, viz., 
$\exp[2N(r_{\rm H})]B(r_{\rm H})=0$. We will also assume that $B(r)$ vanishes 
at 
$r=r_{\rm H}$ and is analytic around $r=r_{\rm H}$. 
However, we will not impose further 
constraints on the function $\exp[2N(r)]$, as it could become singular at 
$r_{\rm H}$, as is easily observed from Eq.(\ref{ein4}).

Multiplying $e^{2N}B$ to Eq.(\ref{ein1}) and taking the limit $r\to r_{\rm H}$,
we find $B'(r_{\rm H})=0$. As $B(r_{\rm H})=B'(r_{\rm H})=0$, at 
$r=r_{\rm H}$, the metric function $B(r)$  can be expanded as:
\begin{eqnarray}\label{app}
B(r)\, := \, B_{\rm H}(r_{\rm H} - r)^2+B_3(r_{\rm H}-r)^3+\cdots.
\end{eqnarray}
Inserting the expansion into Eq.(\ref{ein4}), one finds
\begin{equation}\label{napp}
N'(r)\, :=\, \frac{\beta}{(r_{\rm H}-r)^2}+\frac{\alpha}{(r_{\rm H}-r)}
+{\cal O} [(r_{\rm H}-r)^0],
\end{equation}
where the coefficients $\alpha, \beta$ are given by:
\begin{eqnarray}
\alpha&\equiv&\frac{p+2}{p+1}
          \left[1-\frac{4|\Lambda|}{(p+2)(p+n-1)B_{\rm H}}\right], 
 \label{alph}\\
\beta&\equiv&\frac{-1}{p+1}\left[\frac{8\pi G_D-n+2}{r_{\rm H}^2}
          -\frac{2|\Lambda|}{p+n-1}\right]. \label{beta}
\nonumber
\end{eqnarray}
Apparently, Eq.(\ref{napp}) indicates that, as $r \rightarrow r_{\rm H}$,
$N'(r)$ is divergent quadratically and hence $e^{2N}B$ is divergent 
exponentially. However, Eq.(\ref{ein1}) forces the coefficient $\beta$ to be 
zero, as can be seen straightforwardly from substituting Eqs.(\ref{app}) and 
(\ref{napp}) into Eq.(\ref{ein1}). Thus, $N(r)$ is at most logarithmically 
divergent at $r = r_{\rm H}$. As $\beta=0$, from Eq.(\ref{beta}), we also
obtain: 
\begin{equation}
r_{\rm H}^2 = \frac{(p+n-1)(8\pi G_D-n+2)} {2|\Lambda|} 
\label{rh}.  
\end{equation}
This is one of the central relations we shall be using later for stabilizing
large mass hierarchy. Using the series expansions given by Eqs.(\ref{app}) and 
(\ref{napp}), we obtain from Eqs.(\ref{ein1}) and (\ref{ein3}) that: 
\begin{equation}\label{bha}
B_{\rm H} (1-\alpha)^2=\frac{2|\Lambda|}{(p+1)(p+n-1)},
\end{equation}
and, from simultaneous solution of Eqs.(\ref{bha}) and (\ref{alph}), 
\begin{eqnarray}
\left( 1-\alpha \right) ={1 \over 4} \left[1+\sqrt{(p+9)/(p+1)} \right].
\nonumber
\end{eqnarray}
We have dropped the negative root of $(1-\alpha)$, as, in that case, 
the norm of $\partial_t$ will diverge at $r = r_{\rm H}$, leading to a 
contradiction with our starting assumption that $r_{\rm H}$ is the $p$-brane
horizon. Then, near the horizon, the spacetime metric Eq.(\ref{metric1}) has
the form: 
\begin{equation}\label{nmetric}
ds^2\, :=\, B_{\rm H}[\sigma(r-r_{\rm H})]^{2(1-\alpha)}\, 
\bar{g}_{\mu\nu}dx^\mu dx^\nu+
            \frac{dr^2}{B_{\rm H}(r-r_{\rm H})^2}+r_{\rm H}^2 
d\Omega_{n-1}^2.
\end{equation}
Here, $\sigma$ is a signature factor taking value $-1$ in the interior region 
($r<r_{\rm H}$) and $1$ in the exterior region ($r>r_{\rm H}$) 
\footnote{Note that, so far, we have examined the near-horizon region from the 
inside 
toward $r_{\rm H}$ 
and hence obtained only the interior metric ($\sigma=-1$). Had
we done the analysis from the outside region, we would then obtain 
the exterior metric ($\sigma=1$).}.
It is easy to confirm that $r=r_{\rm H}$ is a degenerate horizon of the 
timelike Killing vector $\partial_t$: $\partial_t$ becomes null at $r_{\rm H}$ 
and the surface gravity is zero. 
The surface at $r=r_{\rm H}$ is also the Cauchy horizon of 
a $(p+2)$-dimensional anti-de Sitter space. Hence, the spacetime Eq.(\ref{nmetric}) is simply the near horizon geometry of an extreme black $p$-brane, which 
is typically of the form $AdS_{p+2}\times {\bf S}_{n-1}$. 
The curvature radii of $AdS_{p+2}$ and ${\bf S}_{n-1}$ are given by 
$k^{-1}\equiv 1/\sqrt{(1-\alpha)^2B_{\rm H}}$ and $r_{\rm H}$, respectively. 

To see the spacetime structure, we find it convenient to reexpress 
Eq.(\ref{nmetric}) in the Poincar\'e coordinates by changing the radial 
variable as $\exp(-k\chi)\equiv\sqrt{B_H}[\sigma(r-r_{\rm H})]^{1-\alpha}$:
\begin{equation}\label{grg}
ds^2\, := \, \exp\left(-2k\chi\right)\bar{g}_{\mu\nu}dx^\mu dx^\nu
            +d\chi^2+r_{\rm H}^2 d\Omega_{n-1}^2,
\end{equation}
where the curvature scale $k$ of the $AdS_{p+2}$ space is defined by 
$k\equiv (1-\alpha) \sqrt{B_{\rm H}}\sim\sqrt{|\Lambda|}$. 
For the interior metric ($\sigma=-1$), $\chi$ runs from a finite value to 
$\infty$ (at $r=r_{\rm H}$). For the exterior solution ($\sigma=1$), $\chi$ 
runs between $\infty$ (at $r=r_{\rm H}$) and $-\infty$ (at $r=\infty$), but 
$\chi$ ought to be truncated at a finite distance as, at sufficiently large $r$,
the near-horizon geometry is replaced by an asymptotic spacetime 
Eq.(\ref{amet}). The near-horizon geometry Eq.(\ref{grg}) coincides with the 
cigar-like spacetime geometry with an exponentially damping warp-factor 
discovered by \cite{Gre} and \cite{Vil}.
Hence, among the solutions found in \cite{Gre} and \cite{Vil}, the ones 
relevant for the RS scenario would be interpreted quite naturally as the 
near-horizon geometry of global black $p$-branes. Moreover, we also find that
the two solutions, Eqs.(\ref{amet}) and (\ref{nmetric}), which were 
apparently treated disjointly in \cite{Gre} and \cite{Vil}, can be matched 
to each other. 

On the other hand, in the region between the $p$-brane core and the horizon, 
where the cosmological constant is negligible compared to the field energy 
density \footnote{If $\Lambda$ is much smaller than the fundamental scale
$M_*^2$, this region overwhelms the near-horizon region.}, the spacetime 
geometry depends on the transverse space dimensions. When $n=2$, the geometry 
of this region resembles that of the Cohen-Kaplan solution of Ref.\cite{CK}. 
If $n\geq3$, the geometry of this region would become similar to that of the 
global monopole solution \cite{Vil}. 

While we have not resolved yet whether the geometries obtained in separated 
regions can be connected mutually or not, painstaking numerical calculations
indicate that the black $p$-brane solution does exist. 
Putting the above results together, the spacetime geometry of the global 
$p$-brane is illustrated in Figure 1. 
\begin{figure}[htb]
\begin{center}
\epsfxsize=6in\epsfysize=3in\leavevmode\epsfbox{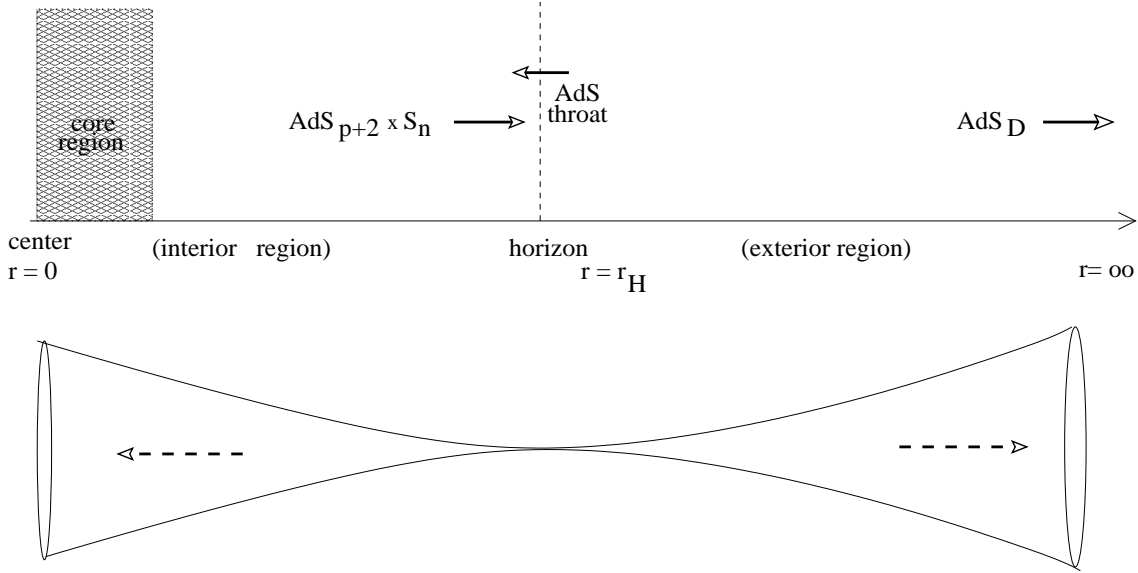}
\end{center}
\caption{(a) Spacetime geometry of the global $p$-brane, (b) Corresponding
warped geometry
in Poincar\'e coordinate.}
\label{cartoon}
\end{figure} 

%%%%%%%%%%%%%%%%%%%%%%%%%%%%%%%%%%%%%%%%%%%%%%%%%%%%%%%%%%%%%%%%%%%%%%%%%%%%%%
\section{Effective Theory of The New Brane-World Gravity}
%%%%%%%%%%%%%%%%%%%%%%%%%%%%%%%%%%%%%%%%%%%%%%%%%%%%%%%%%%%%%%%%%%%%%%%%%%%%%%

Having understood the spacetime geometry generated by the global $p$-brane,
we now proceed to a new brane-world scenario. The first step would 
be identifying a suitable slice of the spacetime geometry, where an 
effective $(p+1)$-dimensional gravity may arise at low-energy. 

The `exterior region' by itself does not seem to provide a suitable framework
for the brane-world scenario, as, in one direction, one will reach the 
$(p+2)$-dimensional AdS horizon while, in the other direction, one will reach 
the $D$-dimensional AdS spacetime. On the other hand, the interior region 
possesses all the features necessary for realizing the RS scenario. In one 
direction, it asymptotes to the AdS throat ($AdS_{p+2}\times {\bf S}_{n-1}$). 
As such, the central region surrounded by the AdS region looks like a 
{\sl one-sided} RS domain-wall embedded in $(p+2)$-dimensional AdS spacetime, 
viz. a smooth Planck-brane in the RS scenario.  

In the RS scenario, as is investigated extensively, the standard 
$(p+1)$-dimensional Newtonian gravity is reproduced in the brane-world 
{\sl provided} a massless $(p+1)$-dimensional graviton is localized on the 
$p$-brane. In the new brane-world scenario, 
existence of the massless graviton is evident, as the field equations 
Eqs.(\ref{ein1})-(\ref{ein4}) permits always a solution with a general Ricci 
flat metric $\bar{g}_{\mu\nu}(x)$ -- the massless graviton is simply the 
standard travelling gravitational wave of the linearized $(p+1)$-dimensional 
vacuum Einstein's field equation. Then, 
the massless $(p+1)$-dimensional graviton is 
localized on the $p$-brane {\sl provided} the $(p+1)$-dimensional Planck scale 
$M_{\rm Pl}$ is finite. From the $(p+1)$-dimensional effective 
action, one identifies the Planck scale with:
\begin{eqnarray}\label{pla}
M_{\rm Pl}^{p-1}= M_*^{D-2}\int dy^n \sqrt{g_D}~g^{00} 
            \sim M_*^{p-1+n}\, r_{\rm H}^n .   
\end{eqnarray}
The relation Eq.(\ref{pla}) shows that the 
$(p+1)$-dimensional Planck scale is determined by the fundamental scale 
$M_*$ and the horizon size $r_{\rm H}$ quite analogous to the usual 
Kaluza-Klein theories {\sl except} that now the size of the compact extra
dimension in the Kaluza-Klein theories is played by the size of the horizon 
$r_{\rm H}$. Hence, in the new brane-world scenario, we propose to interpret
the horizon size $r_{\rm H}$ as an effective size of the noncompact 
$n$-dimensional space, even though the interior region extends to infinity. 
The $n$ extra dimensions turn into a direct product of an 
$(n-1)$-dimensional sphere ${\bf S}_{n-1}$ of a fixed radius $r_{\rm H}$ times
an infinitely extended radial direction. Despite being noncompact, due to 
the warped spacetime geometry, the radial direction yields a finite 
effective size $\sim r_{\rm H}$.

Indeed, to an observer living outside the horizon (the exterior region), 
the above interpretation seems correct. The interior region occupies only a 
finite part 
of the higher dimensional transverse space of volume $\sim r_{\rm H}^n$ and 
the apparent infinite extension of the interior region is simply a result of 
the warping of the extra dimensions by the gravity produced by the $p$-brane 
itself. Later, we will support the interpretation further with analysis of 
the effective $(p+1)$-dimensional gravity.

To see the behavior of $(p+1)$-dimensional gravity on the global $p$-brane, 
let us study small perturbations of the spacetime metric: replacing 
$\bar{g}_{\mu\nu}(x)$ 
with $\eta_{\mu\nu}+h_{\mu\nu}(x,z)$ in Eq.(\ref{metric1}). Imposing the 
so-called Randall-Sundrum gauge, 
one can easily find linearized field equation for 
$h_{\mu\nu}$. For a general metric of the form in Eq.(\ref{metric1}), 
after a change of variables  $\xi=\int^r\sqrt{-g^{00}(r')g_{rr}(r')}~dr'$, 
$ \widetilde{h_{\mu\nu}}= K~h_{\mu\nu}$ and a separation of variables
$\widetilde{h_{\mu \nu}}(\xi,\Omega)=\epsilon_{\mu\nu}e^{ip\cdot x}
{\cal R}_{m\ell}(\xi)
Y_\ell(\Omega)$, where $\epsilon_{\mu \nu}$ is the polarization tensor 
and $Y_\ell (\Omega)$ is the $n$-dimensional spherical harmonics, 
it can be shown that the linearized field equation can 
always be expressed as an `analog' non-relativistic Schr\"{o}dinger 
equation: 
\begin{equation}\label{sch}
\left[-\frac{\partial^2}{\partial\xi^2}+V_{\rm eff}(\xi)\right] 
{\cal R}_{m\ell}(\xi) =m^2 {\cal R}_{m\ell}(\xi),
\end{equation}
in which
\begin{equation}\label{efp}
V_{\rm eff}(\xi)\equiv\frac{K''(\xi)}{K(\xi)}
+\ell(\ell+n-2)\frac{g_{00}(\xi)}{r(\xi)^2}.
\end{equation}
Here, $K\equiv r^{(n-1)/2}g_{00}^{p/4}$  and $m~(=\sqrt{-p\cdot p})$ denotes 
the $(p+1)$-dimensional Kaluza-Klein mass of the continuum modes. 
The zero-mode wave function, with $m=\ell=0$, is easily identified with 
$R_{00}(\xi)=K(\xi)$. All the low-energy physics in the brane-world can then
be analyzed qualitatively from the structure of the effective potential
$V_{\rm eff}(\xi)$. 

Consider a point particle of mass $m^*$ localized on the global $p$-brane. 
Then, the Newtonian potential is calculated straightforwardly as:
\begin{eqnarray}
U(|{\bf x}|) = -G_{p+1}\frac{m^*}{|{\bf x}|^{p-3}}
-\frac{m^*}{M_*^{n+2}}\sum_{\ell}
\int_{m\neq0}dm \, \rho(m) \, |{\cal R}_{m\ell}(0)|^2~
\frac{e^{-m|\bf{x}|}}{|{\bf x}|^{p-3}}. \label{ngr}
\end{eqnarray}
Here, $\rho (m) = m^{\delta}$ denotes the density of states for the continuum
modes.
If the extra dimensions are noncompact, it follows that $\delta=n-1$, as the 
$n$-dimensional plane-waves yields a density of states $m^{n-1}$ (up to a 
calculable numerical coefficient).
In the present case, $(n-1)$-extra dimensions are compactified to a sphere 
${\bf S}_{n-1}$ of radius $r_{\rm H}$. Thus, the lower partial-wave modes 
($\ell \sim 1$) would behave as plane waves only in the radial direction, 
while higher partial-wave modes ($\ell \gg 1$) would behave as plane waves in 
the entire extra dimensions. This implies that $\delta$ will depend on 
$\ell$ and interpolates $\delta=0$ at small $\ell$ and 
$\delta = n-1$ at large $\ell$.

The right-hand side of Eq.(\ref{rh}) ought to be positive. This condition 
yields $v^2>(n-2)/8\pi~M_*^{D-2}$. While the condition does not give any 
new information for $n=2$, for $n > 2$, the condition sets the symmetry 
breaking scale $v$ to be of the oder of the fundamental scale $M_*$.
On the other hand, in order to trust the global $p$-brane profile, 
$v^{2/D-2}$ ought to be smaller than the fundamental scale: $v^{2/D-2}<M_*$. 
Hence, we will be assuming that $(n-2)/8\pi<v^2/M_*^{D-2}<1$. We will also 
assume that $v^{2/D-2}\sim M_*$ for $n=2$, but this seems valid for 
naturalness reasons. The horizon size $r_{\rm H}$ and the anti-de Sitter 
curvature 
radius $k$ are then determined by the bulk cosmological constant $\Lambda$, 
viz, $k^2\sim r_{\rm H}^{-2}\sim |\Lambda|$. 
In models with RS-type scenario, the bulk 
cosmological constant $\Lambda$ had always been identified with the fundamental
scale.
However, given our ignorance concerning the cosmological constant problem, 
there is a priori no reason to expect that $\Lambda$ is of the same order as 
the fundamental scale $M_*^2$. Thus, we shall be treating $\Lambda$ simply as 
an input parameter. The only experimental input is that, from the present-day
gravity measurement, $r_{\rm H}$ ought to be smaller than $\sim 1$mm if it 
is to be interpreted as a compactification scale.  
In what follows, we will consider two limiting values of $\Lambda$:
$|\Lambda|\sim M_*^2\sim 
M_{\rm Pl}^2$ and $|\Lambda|\sim10^{6-60/n}{\rm GeV}^2\ll M_*^2\sim m_{\rm 
EW}^2$ for $p=3$. 
We will call the first limit as Randall-Sundrum (RS) limit and the second 
as large extra dimension (ADD) limit and explore physics in each of them 
in detail. 

%%%%%%%%%%%%%%%%%%%%%%%%%%%%%%%%%%%%%%%%%%%%%%%%%%%%%%%%%%%%%%%%%%%%%%%%%%%%%
\subsection{Randall-Sundrum Limit}
%%%%%%%%%%%%%%%%%%%%%%%%%%%%%%%%%%%%%%%%%%%%%%%%%%%%%%%%%%%%%%%%%%%%%%%%%%%%%

In the RS limit, the interior region is well approximated by the near-horizon
geometry Eq.(\ref{grg}), as both the core radius 
$r_c\sim(\sqrt{\lambda}v)^{-1/2}$ and the horizon size $r_{\rm H}$ are of 
the order of the fundamental scale. As such, $8\pi G_D/r^2$ and $|\Lambda|$ 
are of comparable magnitude over the entire `interior region'. 
The curvature radii of $AdS_{p+2}$ and ${\bf S}_{n-1}$ are of the order of 
the Planck scale. Thus, at low-energy below the Planck scale, the extra space  
is reduced effectively to a one-dimensional space. Consequently, the 
global $p$-brane core looks like a $p$-dimensional domain-wall embedded in 
an $AdS_{p+2}$ bulk spacetime. In the `thin-wall approximation' limit, 
the situation is essentially the same as that of the RS scenario. 
After a change of variable $k\xi+1=\exp(k\chi)$, one obtains 
the effective potential $V_{\rm eff}(\xi)$ as
\begin{equation}\label{spo}
V_{\rm eff}(\xi)=\left(\frac{p(p+2)k^2}{4} +\frac{\ell(\ell+n-2)}
{r_{\rm H}^2}\right) \frac{1}{(k\xi+1)^2}.
\end{equation}
The `analog' 
Schr\"{o}dinger equation Eq.(\ref{sch}) allows a normalized
zero-mode ${\cal R}_{00}\sim r_{\rm H}^{-n/2}(k\xi+1)^{-p/2}$ 
and a continuum of 
Kaluza-Klein modes given by a linear combination of Bessel functions 
$
{\cal R}_{m\ell}(\xi)=(\xi+1/k)^{1/2}\left[a_{m\ell}Y_\nu\big(m(\xi+1/k)\big) 
+b_{m\ell}J_\nu\big(m(\xi+1/k)\big)\right],$ 
where $a_{m\ell}$ and $b_{m\ell}$ are $m$-dependent coefficients, and
$\nu\equiv\frac{p+1}{2}\sqrt{1+4\ell(\ell+n-2)/(p+1)^2k^2r_{\rm H}^2}$.
As anticipated, for $\ell=0$ partial wave, the potential and the 
wavefunction are precisely the same as those of the original RS model. 
A difference from the RS model arises for massive Kaluza-Klein
modes with $\ell\neq0$ extended in $AdS_{p+2}$ --- they originate from the 
Kaluza-Klein reduction of the fundamental $D$-dimensional bulk fields on 
${\bf S}_{n-1}$. Consequently, they exhibit a discrete mass spectrum 
proportional to $\ell(\ell+n-2)/r_{\rm H}^2$.

In order to evaluate the sub-leading contributions due to the Kaluza-Klein
modes to the $(p+1)$-dimensional static gravitational potential, one will 
need to estimate ${\cal R}_{m\ell}(\xi = 0)$ for
$m\ll M_*$. As the $p$-brane core region is extended over the scale 
$r_c$ in the transverse space, it is difficult to impose a suitable
boundary condition, in contrast to the original RS model. However, in case 
the central region of the Schr\"{o}dinger potential is localized within the 
AdS scale $1/k$, the sub-leading contribution would be identical to those of 
the original RS limit (The same argument has been put forward earlier
by \cite{CEHS}.).

Adopting the same procedure as in \cite{CEHS}, we obtain the value of the 
radial wavefunction on the brane at $\xi = 0$ at leading order in small $m$
expansion as ${\cal R}_{m\ell}(\xi\simeq 0)\sim(m/k)^{\nu-3/2}$ for $m\ll k$. 
It is now straightforward to calculate the static gravitational potential 
at long distances $|{\bf x}|\gg 1/M_{\rm Pl}$ generated by a point particle of 
mass $m^*$ located on the $p$-brane:
\begin{eqnarray}
U(|{\bf x}|)&\sim&-G_{p+1}\frac{m^*}{|{\bf x}|^{p-2}}\left[1
+\left(\frac{1}{k|{\bf x}|} \right)^{p-1}+\sum_{\ell\neq0}\left(
\frac{1}{k|{\bf x}|}\right)^{\delta+2\nu-2} \right] . \label{npot}
\end{eqnarray}
The leading term originating from the massless bound-state graviton 
contribution is the usual $(p+1)$-dimensional Newtonian potential. 
The second correction term 
coming from continuum modes with $\ell=0$ is suppressed enormously for $k$ of 
order the $(p+1)$-dimensional Planck scale and the distance scale the gravity
is tested. 
The correction grows like $1/r^{p-1}$ rather than $1/r^{n+1}$, unlike those 
expected for the RS-type models with $n$ extra 
dimensions \cite{ADDK,shap,Oda}. The difference from the RS scenario arises 
only through the third term, which originates 
from the contribution of modes with 
$\ell\geq1$ and was absent in the original RS model. The information of the 
extra $(n-1)$ dimensions  is reflected through this term only. It, however, is 
more strongly suppressed than the second term for large $\ell$. 
Thus, the low-energy processes on the brane are indistinguishable from 
those in the original RS scenario.

%%%%%%%%%%%%%%%%%%%%%%%%%%%%%%%%%%%%%%%%%%%%%%%%%%%%%%%%%%%%%%%%%%%%%%%%%%%%%%
\subsection{Large Extra-Dimension Limit}
%%%%%%%%%%%%%%%%%%%%%%%%%%%%%%%%%%%%%%%%%%%%%%%%%%%%%%%%%%%%%%%%%%%%%%%%%%%%%%

Consider next the case where the cosmological constant $\Lambda$ is 
hierarchically smaller than the fundamental mass scale $M_*$. In this situation, the physics will be quite analogous to that of the ADD scenario \cite{ADD}. 
Adopting the results of \cite{ADD}, in our new brane-world scenario, 
phenomenologically acceptable size of the horizon and the cosmological
constant are $r_{\rm H}\sim 10^{30/n-17}$cm and $|\Lambda|\sim 
10^{6-60/n}{\rm GeV}^2$, respectively, with $M_*\sim m_{\rm EW}\sim10^3$GeV 
and $M_{\rm Pl}\sim10^{19}$GeV for $p=3$. 
In this case, in the transverse space, 
the core region and the $AdS_{p+2}$ throat region are squeezed into small 
portions and, in between the two, most portion is occupied by an intermediate 
region characterized by $8\pi G_D/r^2\gg|\Lambda|$. The proper size of the 
intermediate region $\xi_{\rm I}$ is close to the horizon size: 
$\xi_{\rm I} \sim r_{\rm H}$. As such, depending on details of the geometry
of this region, the functional form of the `analog' Schr\"{o}dinger potential 
will be modified from Eq.(\ref{spo}).

In fact, as mentioned earlier, the geometry of the intermediate region can be 
approximated by that of the Cohen-Kaplan solution \cite{CK} for $n=2$ and 
those of global monopoles \cite{Vil} for $n \geq3$. Let us analyze them in
more detail. For $n=2$, the `analog' Schr\"{o}dinger potential is attractive 
near the core and shoots to a negative infinity at the singularity \cite{CK}. 
This potential ought to be excised at a point near the singularity and then
glued smoothly to a potential generated by the $AdS_{p+2}$ throat region, 
Eq.(\ref{spo}). As the potential takes a negative value in this region, the 
continuum Kaluza-Klein modes are separated by a potential barrier whose peak
is located at the $AdS_{p+2}$ throat region and height is given by $\sim k^2$. 
Furthermore, the potential extends over the $AdS_{p+2}$ curvature scale, $1/k$.
As discussed in the previous subsection, in case the Schr\"{o}dinger potential 
is localized within the scale $1/k$, the effect of light Kaluza-Klein 
continuum modes (whose mass is ranged $m\ll k$) turns out identical to that 
in the RS limit. Hence, the subleading contribution to the long-range static
gravitational potential should be the same as in the RS limit, independent of
details of the potential profile. As a result, we will obtain the 
static gravitational potential with corrections of the same form 
as Eq.(\ref{npot}) (at a long distance, $|{\bf x}|\gg r_{\rm H}\sim k^{-1}$).
The only difference is that the horizon size $r_{\rm H}$ 
is now macroscopically large.  
The continuum Kaluza-Klein modes with masses $m\gg k$ would pass over the 
potential and hence are unsuppressed at the core: ${\cal R}_{m\ell}(0)\sim1$. 
Hence, at a short distance $|{\bf x}|\ll r_{\rm H} \sim k^{-1}$, the static
gravitational potential is just $(p+3)$-dimensional:
\begin{equation}
U(|{\bf x}|)\sim \frac{1}{M_*^{p+1}}\frac{m^*}{|{\bf x}|^p}.
\end{equation}

When $n \geq3$, the intermediate region ($\xi\lsim r_{\rm H}$) is approximated 
by the global monopole geometry \cite{Vil}:
\begin{equation}
ds^2=\eta_{\mu\nu}dx^\mu dx^\nu +d\xi^2+\gamma^2 \xi^2d\Omega_{n-1}^2, 
\end{equation}
where $\gamma^2\equiv 1-8\pi G_D/(n-2)$ represents the solid angle deficit 
in extra dimensions. In this region, the `analog' Schr\"{o}dinger potential 
is repulsive of the form:  
\begin{equation}
V_{\rm 
eff}(\xi)= \left[\frac{(n-1)(n-3)}{4}+\frac{\ell(\ell+n-2)}{\gamma^2}\right]
\frac{1}{\xi^2} ,
\label{potential}
\end{equation}
The potential in the $AdS_{p+2}$ region can be obtained from replacing $\xi$ 
by $\xi-\xi_I$ in Eq.(\ref{spo}). 

For $n=3$, Eq.(\ref{potential}) indicates that, in the intermediate region, 
the repulsive potential term vanishes identically and hence imposes no 
further suppression for the Kaluza-Klein continuum modes. As such, the 
physics on the $p$-brane would be the same as that of the ADD scenario with 
$n=3$. We thus find that, for $n=2,3$, the effective low-energy gravity 
behaves as expected in a world with $n$-extra compactified dimensions 
of radius $r_{\rm H}$.  

For $n\geq4$, the physics on the $p$-brane will be quite different from that
of $n=2,3$, mainly due to the repulsive potential in the intermediate region. 
As $\xi_{\rm I}\sim 1/k$, the potentials in the two regions are matched 
smoothly at $\xi_{\rm I}$ and the potential increases monotonically up to 
$\sim M_*^2$ as approached to the core region 
$r_c\sim (\sqrt{\lambda}v)^{-1}\sim M_*^{-1}$. In this region, the 
Kaluza-Klein continuum modes are given by combinations of Bessel functions 
$\sqrt{\xi}\left[A_{m\ell}J_{\upsilon}(m\xi) 
+B_{m\ell}Y_{\upsilon}(m\xi)\right]$ where $\upsilon\equiv \left(
\frac{n-2}{2}\right) \left[1+4\ell(\ell+n-2) /(n-2)^2\gamma^2\right]^{1/2}$.  
The modes with mass $m\lsim M_*$ are suppressed at the core and, at
leading order in $m$, is given by ${\cal R}_{m\ell}(0)
\sim(m/M_*)^{\upsilon-3/2}$.
As such, the static gravitational potential generated by a point particle 
with mass $m^*$ at distances $|{\bf x}|\gg M_*^{-1}$ located on the $p$-brane
is simply obtained by replacing $k,~ \nu$ with $M_*,~\upsilon$ in 
Eq.(\ref{npot}). 
%On the other hand, for distances $|{\bf x}|\ll M_*^{-1}\ll 
%r_{\rm H}$, the gravity becomes fully $D$-dimensional. 

Actually, this regime 
is phenomenologically interesting as the physics on the $p$-brane is different
from those of the RS- or the ADD-limits. For the RS scenario, by naturalness,
the fundamental scale and the bulk cosmological constants are assumed to be 
of the same order as the Planck scale $M_{\rm Pl}$. Moreover, the gravity in 
the brane-world is not significantly modified over the 33 orders of magnitude 
between $\sim1$cm and the Planck length $M_{\rm Pl}^{-1}\sim10^{-33}$cm.
For the  ADD scenario, the fundamental scale may be taken of the order of the
electro-weak scale $m_{\rm EW}$ but, to obtain the correct Planck scale, 
the compactification size ought to be macroscopically 
large, leading to very light 
continuum modes. Consequently, the gravity in the brane-world changes from 
four- to higher-dimensional gravity around the compactification scale, 
much larger than the electro-weak scale $1/ m_{\rm EW}$. However, in the last
case with $n\geq4$, the continuum modes lighter than the fundamental mass 
scale 
$\sim m_{\rm EW}$ are suppressed and the gravity in the brane-world remains 
$(p+1)$-dimensional down to the distance $\sim m_{\rm EW}^{-1}$.
Nor do very light moduli fields associated with the large extra dimensions 
${\bf S}_{n-1}$ get problematic, as they are strongly suppressed by the 
potential barrier of height $M_*^2\sim m_{\rm EW}^2$. 

%%%%%%%%%%%%%%%%%%%%%%%%%%%%%%%%%%%%%%%%%%%%%%%%%%%%%%%%%%%%%%%%%%%%%%%
\section{Large Mass Hierarchy from Large Horizon Size}
%%%%%%%%%%%%%%%%%%%%%%%%%%%%%%%%%%%%%%%%%%%%%%%%%%%%%%%%%%%%%%%%%%%%%%%

So far, we have obtained a new brane-world, whose spirit is similar to 
ADD scenario for $n=2,3$ but distinct for $n\geq4$. In all cases, 
the large mass hierarchy is explained by a large horizon size $r_{\rm H}$.
Recall that a large horizon size is provided by a small bulk cosmological
constant. As such, the hierarchy problem is now transmuted into a large 
disparity between the fundamental scale and the small bulk cosmological 
constant. 
This hierarchy may be stable in that a small change of the bulk cosmological 
constant impart little influence on the low-energy dynamics on the brane-world
as expected in the conventional large extra dimension scenario. 
Then, remaining question is whether the mass hierarchy can be set large by 
underlying dynamics. Generically, this is impossible as
radiative corrections would drive the bulk cosmological constant to the
value set by the fundamental mass scale, possibly except supersymmetric
bulk theory.    

We now point out a novel mechanism for maintaining a large mass hierarchy
in the new brane-world {\sl without} fine-tuning of various scales involved. 
We begin with an observation that, for the global
$p$-branes, the fundamental scale can be separated from the horizon scale 
parametrically large. Recall that, for the extremal $p$-brane, the horizon
size $r_{\rm H}$ is proportional to the global charge $Q$ carried by the 
brane. The only remaining question is whether the $p$-brane remains stable
once its global charge $Q$ is cranked up to a large value. Note that, so
far, we have restricted our considerations to the $\vert Q \vert = 1$ case, 
as, for $n \ge 3$ (global (hyper)-monopoles), they are the only ones 
compatible with spherical symmetry and regularity at the origin.
On the other hand, for $n=2$, a rotationally symmetric solution with an 
arbitrarily large $Q$ is possible!

The $n=2$ configuration with a large winding number $Q$ is obtained simply by 
replacing $8\pi G_D$ in the unit charge configuration with $8\pi G_D Q^2$. 
Then, from Eq.(\ref{rh}), we see that the horizon size is given, in terms of 
the unrescaled parameters, by
\begin{equation}
r_{\rm H}^2=\frac{4\pi v^2(p+1)}{M_*^{p+1}|\Lambda|}~Q^2,
\end{equation}
whereas the curvature of the $AdS_{p+2}$ is unchanged and is determined 
solely by the bulk cosmological constant. We further assume most conservatively
that there exists only one physical scale, viz. all scales are set by the 
fundamental scale. In this case, the horizon size is simply estimated as: 
\bea
r_{\rm H}\sim {1 \over M_*} \vert Q \vert.
\nonumber
\eea  
Putting $M_{*}\sim m_{\rm EW}\sim 10^3$GeV and that $r_{\rm H}$ and $\vert Q
\vert$ are chosen to reproduce the observed 4D Planck scale 
$M_{\rm Pl}\sim 10^{19}$GeV yields, for $p=3$,  we find $Q\sim 10^{16}$. 
Hence, an extremal 3-brane with large global charge, $Q\sim 10^{16}$, provides 
for a dynamical determination of the large hierarchy between the Planck scale 
and the electro-weak scale despite the microscopic theory contains only one
physical scale. Moreover, the large compactification size is now stabilized via 
{\sl topological mechanism}, as the global charge is discrete and conserved.
 
The only drawback with the above mechanism is that, the $p$-brane thickness
$r_c$ 
may become much larger than $1/ m_{\rm EW}$, as the thickness is proportional
to the global charge $Q$. This does not look like the world we live in. 
Nevertheless, we could avoid this drawback, {\sl provided} the thickness 
$r_c$ is much smaller than the horizon size, viz. $m_{\rm EW}^{-1}\ll 
r_c\ll r_{\rm H}$. In this case, one obvious strategy is to place the 
`Standard Model'-brane not at the center but somewhere between $r_c$
and $r_{\rm H}$, similar in spirit to Ref.\cite{LR,ADDK}. The `Standard-Model'
brane should be much thinner than the centrally located 3-brane of charge $Q$, 
viz. a 3-brane with a small topological charge. 

There are two options of the `Standard-Model'-brane location, $r_{\rm SM}$. 
One option is to put it in the $AdS_5$ region: 
$r_{\rm SM} \sim r_{\rm H}$, as in Ref.\cite{LR,ADDK}. In this case, as the 
graviton amplitude is exponentially suppressed, one needs to fine-tune the 
location of the `Standard Model'-brane with high precision in order to 
obtain the correct ratio between the Planck and the electroweak scales. 
Another option is to put it somewhere between the core and the $AdS_5$. 
In this region, the graviton amplitude falls off slowly and no fine-tuning
is required for the the location of the `Standard Model'-brane.
The physics on the `Standard-Model' brane would be the same as in the 
conventional large extra dimension scenario. In the case of a large topological 
charge, the behavior of the continuum modes is more interesting. The horizon 
size is much larger than the fundamental length scale, while the hight of the 
`analog' Schr\"{o}dinger potential is $\sim m_{\rm EW}^2$. 
The central part of the potential then is much larger than the 
curvature length scale; $\xi_{\rm I}\sim r_{\rm H}\gg m_{\rm EW}^{-1}$. 
Thus, we cannot directly apply the procedure 
of Ref.\cite{CEHS}, as we did in previous subsections to see the limiting 
behavior of the continuum modes in the central region of the analog 
Schr\"{o}dinger potential.
This regime seems to mimic the large box limit of Ref.\cite{LMW}, which 
consists of a slice of 5D flat Minkowski space glued 
between two 5D AdS regions. 
In that setup, when the size of the box is large compared to the
anti-de Sitter 
scale, there are resonant modes with enhanced support inside the box.
Even though there is a continuum of bulk modes with masses near the resonant
mass, their net contribution mimics a single normalizable mode with this 
resonant mass. Thus the details of the anti-de Sitter 
region are suppressed in this 
setup. That is, the leading order corrections to the Newtonian potential are
identical to those as if we were considering a compactification of flat
five-dimensional spacetime on ${\bf S}^1/{\bf Z}_2$ orbifold. 
The regime with large winding number
clearly resembles this large box limit of Ref.\cite{LMW} and corresponds to 
the generalization of their setup \cite{LMW} to a spacetime with more 
than five dimensions. The interior
region has a finite portion of a space, even though it is not a flat space, 
surrounded by an anti-de Sitter region. Following the works of Ref.\cite{LMW}, we easily 
deduce that there ought to be resonant modes with masses $m_\ell \sim 
\ell/r_{\rm H}$, where $\ell=1,2,3,...$. Hence, the low energy physics on the 
brane imitates the conventional large extra dimension scenario. 

The phenomenology on the `Standard Model' brane located in the region 
between the core and the anti-de Sitter region resembles closely 
that of the large 
extra dimension scenario except that the size of the large extra dimensions 
is provided by the large topological charge carried by the global black 
3-brane and is stabilized via the topological charge conservation. 
Stated differently, a global black brane with a large topological charge can 
determine dynamically a stable hierarchy between the 
four-dimensional Planck scale and the electroweak scale in a higher-dimensional
theory with a single fundamental scale.

%%%%%%%%%%%%%%%%%%%%%%%%%%%%%%%%%%%%%%%%%%%%%%%%%%%%%%%%%%%%%%%%%%%%%%
\section{Thinning Out Cosmological Constant via Hawking Radiation}
%%%%%%%%%%%%%%%%%%%%%%%%%%%%%%%%%%%%%%%%%%%%%%%%%%%%%%%%%%%%%%%%%%%%%%

Many interesting attempts for the resolution of the cosmological constant 
problem have been claimed 
within the brane world scenarios \cite{CC}: If our four-dimensional world is 
embedded in a higher dimensional spacetime, changes of the vacuum energy
of the brane, the brane tension, may affect the curvature in the extra 
dimensions only, retaining 
a Poincar\'{e} invariant four-dimensional worldvolume. 
These scenarios have undesirable features, such as the presence of naked 
singularity parallel to the brane, or the necessity of very specific form of 
the couplings in the effective action of the theory. 

In this section, we would like to propose one intriguing piece of physics 
related to the cosmological constant -- relaxation of the cosmological 
constant of the brane world via Hawking radiation. Till now, we have considered 
only the extremal black brane, of which world volume is Ricci-flat.  
For such extremal black branes, the Hawking temperature is zero because the 
surface gravity vanishes at the horizon. Therefore, there would be no Hawking 
radiation either. 

On the other hand, a non-extremal black brane has nonzero Hawking temperature
and radiates, evolving toward an extremal black brane. 
We take this as a possible mechanism for relaxing the cosmological 
constant in the brane world scenario.  
In general, excitations of the extremal black $p$-brane would convert the 
metric $\eta_{\mu\nu}$ to be either non-Ricci flat $\hat{g}_{\mu \nu} = 
\hat{g}_{\mu \nu}(x)$ such that $\hat{R}(\hat{g})\neq0$, or even to depend on 
the extra dimension coordinates $\hat{g}_{\mu \nu} = \hat{g}_{\mu \nu}(x, y)$. 
The non-extremal black $p$-brane produced so will be equipped with nonzero 
Hawking temperature. Subsequently, the excitations will be diluted via Hawking 
radiation. 
%How would the relaxation process 
%proceed? The Kaluza-Klein continuum modes are essential part of such process 
%in so far as they are nonsingular. In pure anti-de Sitter background,
%they indeed exhibit singular behavior at the horizon\cite{CG}. However, 
%in general background as in the present case, they are nonsingular and hence
%do participate in the Hawking radiation process. 

In this picture, the nonzero vacuum energy of the brane is translated into the 
constant worldvolume curvature $\hat{R}(\hat{g})\neq0$, which would render the 
initially extremal black $p$-brane into a non-extremal one. 
This observation implies that, even though 
the Poincar\'{e} invariance along the black $p$-brane worldvolume direction
is broken due to the change of the vacuum energy density of the brane, 
{\it e.g.}, the quantum corrections to the brane tension, it can be recovered
through a dilution mechanism of the excitation via the Hawking radiation 
process. Consequently, one would expect that no cosmological constant on the 
brane would be generated, at least, at asymptotic future.
In this picture, the observed smallness of the cosmological constant could 
be explained, provided that our world brane is embedded in the interior region 
of a very near-extremal black brane. Moreover, the fact that the entropy 
density of our Universe is extremely small (but nonzero) compared to the 
entropy density usually expected from black hole physics seems to 
support it \footnote{ The present entropy density of our universe is 
$\sim 10^3/{\rm cm}^3$, while for a four-dimensional black hole which has a 
Schwarzschild radius of order 1cm, the entropy is $\sim 10^{66}$.}.
  
The above argument is applicable as well
for other problems associated with the brane world scenario.   
First, the process provides a new solution to the cosmological 
flatness problem. Even though the $p$-brane worldvolume might be bent 
initially, the bending energy ought to be diluted as the black brane Hawking 
radiates. The $p$-brane would evolve again toward the extremal one at 
asymptotic future. 
Second, the new flatness problem\cite{CKR} associated with the approximate 
Lorentz invariance in the $p$-brane worldvolume direction may
also be explained, as the bulk curvature generated by $\hat{g}_{\mu \nu}(x, y)$
could be diluted away via the aforementioned Hawking radiation.
 
%%%%%%%%%%%%%%%%%%%%%%%%%%%%%%%%%%%%%%%%%%%%%%%%%%%%%%%%%%%%%%%%%%%%%%
\section{Summary and Discussion}
%%%%%%%%%%%%%%%%%%%%%%%%%%%%%%%%%%%%%%%%%%%%%%%%%%%%%%%%%%%%%%%%%%%%%%

In this paper, we have put forward a new kind of brane-world scenario. The
scenario is based on the existence of a global black $p$-brane in a
higher-dimensional gravity. It represents a black hole-like topological soliton
and is a $p$-dimensional extended object surrounded by a degenerated horizon.  
The geometry is perfectly regular everywhere. We have found that the interior 
region of such a global black brane possesses all of the features necessary 
for realizing the RS-type brane-world scenario. 
In this picture, the size of the horizon can be interpreted as the effective 
size of $n$ compact extra dimensions in that the Planck scale $M_{\rm Pl}$ is 
determined by the fundamental scale $M_*$ and the horizon size $r_{\rm H}$ 
via the familiar relation $M_{\rm Pl}^{p-1}\sim M_*^{p-1+n}r_{\rm H}^n$.
                                                                               
If the fundamental scale and the bulk cosmological constant are taken to be 
the brane world Planck scale, {\it i.e.,} $M_*\sim |\Lambda|^{1/2}\sim M_{\rm
Pl}$, then the size of the horizon is of the order of the Planck scale, 
$r_{\rm H} \sim M_{\rm Pl}$, 
and the low energy physics is essentially the same as in 
the original RS scenario. That is, the corrections to the Newtonian gravity 
have precisely the same power law behavior as in the RS scenario, despite the 
existence of more than one extra dimensions.   

In case $M_*\sim m_{\rm EW}$, the horizon size should be large 
to reproduce the correct brane world Planck scale. When $n=2,3$, the gravity 
behaves as expected in a world with $n$ extra dimensions compactified to a 
size $r_{\rm H}$. 
On the other hand, for $n \ge 4$, there are significant changes in the 
low-energy physics. 
In this case, the continuum modes with mass $m<m_{\rm EW}$ are 
suppressed and the gravity on the brane is maintained to be $(p+1)$-dimensional
down to distances of $\sim m_{\rm W}^{-1}$, in contrast to the conventional 
large extra dimension scenarios. Moreover, there 
are no light moduli fields associated with the large size of ${\bf S}^{n-1}$ 
as they are strongly suppressed in the central region.
The hierarchy between $M_{\rm Pl}$ and $m_{\rm EW}$ is supported by large size 
of the horizon. However, there still remains a hierarchy between the 
weak scale and the tiny bulk cosmological constant is needed for 
a large horizon size.

Finally, in case the theory has intrinsically only one physical
scale of the weak scale, {\it i.e.,} $M_*\sim|\Lambda|^{1/2}\sim m_{\rm EW}$,  
the large horizon size is provided by a large winding number, 
$r_{\rm H}\sim \vert Q \vert~m_{\rm EW}$, 
without introducing small cosmological constant.
The large hierarchy between $M_{\rm Pl}$ and $m_{\rm EW}$ is now provided
by the large winding number of the global black $p$-brane and is stabilized
via the topological charge conservation.
%Even though the size of the horizon is much larger than the hight of the analog
%Schr\"{o}dinger potential, the low energy physics on a brane world residing 
%in the central region imitates the conventional large extra dimension scenario
%due to the existence of resonant modes with masses $m_l\sim l/r_H$. 

Perhaps the deepest consequence of our new brane world is that it offers a 
natural 
explanation of the large mass hierarchy between the four-dimensional Planck 
scale $M_{\rm Pl}$ and the weak scale $m_{\rm EW}$. The large mass hierarchy is 
translated into the large size of the horizon. The large size of the horizon, 
{\it i.e.,} the small compactification scale is provided with the large 
magnitude of charge carried by the black branes, that is, winding number. 
Hence, the macroscopic compactification size now is a quantity calculated from
the microscopic theory and its stabilization is guaranteed from the charge
conservation. 

This picture also implies that one does not need to compactify all of the
extra-dimensions to get the observed four-dimensional world. From an observer's
viewpoint in the higher-dimensional spacetime, near the horizon of the black 
branes, only the $(n-1)$ extra dimensions are compactified locally into 
${\bf S}^{n-1}$ and the rest one extra dimension is warped. 
The apparent infinite extent of the near horizon
space is simply a consequence of the warped region. Clearly, the interior 
region takes only a finite part of the higher-dimensional transverse space
with volume $\sim r_{\rm H}^n$ and its apparent infinite extent is due to the 
warping by the gravity of the brane itself. Hence, the setup needed for 
a realistic brane world can be obtained simply from 
noncompact higher-dimensional spacetime by means of formation of such 
black-brane.   

\section*{Acknowledgement}
We are grateful to S. Dimopoulos, N. Kaloper and Choonkyu Lee for 
discussions, and I. Antoniadis and N. Arkani-Hamed for correspondences.

\end{document}